\documentclass[conference]{IEEEtran}
\usepackage{graphicx}
\usepackage{amsmath}
\usepackage{lipsum}
\usepackage{amsthm}
\usepackage{amsfonts}
\usepackage{caption}
\usepackage{blkarray}
\usepackage{mathtools}
\usepackage{siunitx}
\usepackage{amssymb}
\usepackage{subcaption}
\usepackage{placeins}
\usepackage{xcolor}
\usepackage{lettrine}
\usepackage{multicol, blindtext}
\usepackage{bbm}
\usepackage{cite}
\usepackage{algorithm}
\usepackage{hyperref}
\usepackage{algpseudocode}
\usepackage{authblk}
\newtheorem{remark}{Remark}
\makeatletter

\newcommand{\Rmnum}[1]{\expandafter\@slowromancap\romannumeral #1@}

\newtheorem{theorem}{Theorem}

\newtheorem{proposition}{Proposition}

\newtheorem{definition}{Definition}
\newtheorem{lemma}{Lemma}

\makeatletter
\def\BState{\State\hskip-\ALG@thistlm}
\makeatother
\ifCLASSINFOpdf
\else
\fi

\begin{document}

\title{Minimizing the Age of Incorrect Information for Real-time Tracking of Markov Remote Sources}


\author{ Saad Kriouile}\author{Mohamad Assaad}
\affil{Laboratoire des Signaux et Syst\`emes, CentraleSup\'elec, Universit\'e Paris-Saclay,  91192 Gif sur Yvette, France}

\maketitle
\newcommand{\HRule}{\rule{\linewidth}{0.5mm}}

\begin{abstract}

The age of Incorrect Information (AoII) has been introduced to address the shortcomings of the standard Age of information metric (AoI) in real-time monitoring applications. In this paper, we consider the problem of monitoring the states of remote sources that evolve according to a Markovian Process. A central scheduler selects at each time slot which sources should send their updates in such a way to minimize the Mean Age of Incorrect Information (MAoII). The difficulty of the problem lies in the fact that the scheduler cannot know the states of the sources before receiving the updates and it has then to optimally balance the exploitation-exploration trade-off. We show that the problem can be modeled as a partially Observable Markov Decision Process Problem framework. We develop a new scheduling scheme based on Whittle’s index policy. The scheduling decision is made by updating a belief value of the states of the sources,   which is to the best of our knowledge has not been considered before in the Age of Information area. To that extent, we proceed by using the Lagrangian Relaxation Approach, and prove that the dual problem has an optimal threshold policy. Building on that, we shown that the problem is indexable and compute the expressions of the Whittle’s indices. Finally, we provide some numerical results to highlight the performance of our derived policy compared to the classical  AoI metric.

\end{abstract}

\section{Introduction}
The notable advance in wireless technology and the availability of low-cost hardware have led to the emergence of real-time monitoring services. In these systems, the monitor needs to know the status of one or multiple processes observed by remote sources. Specifically, the sources send packets that contain the information about the process of interest to the monitor to perform a given task. To that extent, the main goal in these applications is to keep the monitor up to date by receiving the fresh information from different sources. This concept of freshness is captured by the Age of Information (AoI) which is introduced for the first time in \cite{kaul2012real}. Since then, the AoI has become a hot research topic, and a considerable number of research works have been published on the subject  \cite{maatouk2020optimality,hsu2019scheduling,kadota2018scheduling}.  Although this metric quantifies the information time lag at the monitor, it fails to capture the correctness of the information at the monitor side.  Specifically, the evolution of this metric doesn't take into consideration the state of the  information at the monitor side.
This has been confirmed in \cite{jiang2019unified} where the authors establish that minimizing AoI gives a sub-optimal policy in minimizing the status error in remotely estimating Markovian sources.
To deal with this issue, some works propose to minimize the estimation
error or the mean square error \cite{sun2019sampling,kam2018towards}. However, the metrics developed in these works are unable to capture the concept of freshness. In other words, there is no penalty incurred to the monitor or the central entity for being in incorrect state for a long time.

To meet the timeliness requirement in the process estimation framework, the authors  in \cite{maatouk2019age} have designed a new metric dubbed as Age of incorrect information AoII that captures the freshness of the information while taking into account the information content acquired from the transmitter. This metric is adopted in the context where a given source is represented as a process denoted by $X(t)$ and the transmitter send status update to the receiver to inform it about the current state of $X(t)$. Under energy or transmission rate constraint, the transmitter cannot use at each time slot the channel to transmit the packet. In this case, as long as the transmitter is in the idle mode, the monitor keeps the last information which may be in the erroneous state compared to the current state of the process $X(t)$. Denoting by $\hat{X}(t)$ the estimated state in the monitor side, being at the incorrect state or equivalently $\hat{X}(t) \neq X(t)$, is clearly an undesirable situation with regards to the monitor, and therefore a penalty should be paid. The metric developed
by \cite{maatouk2019age}, AoII matches with this notion of penalty. Specifically, unlike AoI, AoII evolves only if the estimated state $\hat{X}(t)$ in the monitor side is different from the real state of the process of interest $X(t)$. While if $X(t)=\hat{X}(t)$, the AoII doesn't evolve. To that extent in \cite{maatouk2019age,maatouk2020age}, the authors consider the problem of minimizing the average AoII in a transmitter-receiver pair scenario where packets are sent over an unreliable channel subject to a transmission rate constraint. They derive the optimal
solution which is of the form threshold-based policy. 
The work in \cite{kam2020age} studies the AoII metric in the simple context of monitoring a symmetric binary information source over a delay system with feedback. The authors proposes a dynamic programming algorithm to compute the optimal sampling policy. 

However, \cite{maatouk2019age,maatouk2020age,kam2020age} assume that the scheduler has a perfect knowledge about the process $X(t)$ at each time slot $t$ and restrict the analysis to one transmitter-receiver pair communication. On the opposite of that, in this paper, we tackle a realistic case in which a scheduler tracks the states of multiple remote sources and selects at each time a subset of them send their updates, in such a way to minimize the Mean Age of Incorrect Information (MAoII). Furthermore, the scheduler does not know the instantaneous state of the remote sources until it receives their updates.
Specifically, our contributions can be summarized as follows:
\begin{itemize}
	\item Since the scheduler cannot know at each time the current states of the  sources before it receives their updates, it cannot know exactly the value of MAoII and has to track/predict its evolution. To that end, we introduce a belief state at the monitor, which can be interpreted as the probability that the state at the monitor side is correct, i.e. $\hat{X}(t)=X(t)$. We then describe how this belief state can be derived and can be used in the development of the scheduling policy. 
	\item We then formulate the MAoII-based scheduling problem and show that it belongs to the family of Restless
	Multi-Armed Bandit (RMAB) problems. The optimal solution of this type of problem is known to be out of reach. To circumvent this difficulty, we develop the low-complex and efficient policy called Whittle's index policy (WIP) using the Lagrangian Relaxation Approach.   

\end{itemize}
\section{System Model}\label{sec:Syst_mod}
\subsection{Network description}\label{subsec:Net_descrip}
We consider in our paper $N_u$ users that generate and send status updates about the process of interest to a central entity over unreliable channels. Time is considered to be discrete
and normalized to the time slot duration.
More specifically, each user $i$ observes an information process of interest $X_i(t)$ and at the request of the monitor, it samples the process $X_i(t)$ and send it to the monitor over an unreliable channel. Based on the last received update, the monitor constructs an estimate of the process, denoted by $\hat{X}_i(t)$.  
Given that the time duration of packet's transmission is one time slot, then if the monitor allows a user $i$ to transmit at time $t$, it receives the value of $X_i(t)$ at time slot $t+1$ in the case where the packet is successfully transmitted. Therefore, it updates the estimate process as $\hat{X}_i(t+1)=X_i(t)$.
In any other case, namely when the user $i$ is not authorized to transmit or when the packet is unsuccessfully transmitted, the monitor keeps the same value at time slot $t$, specifically $\hat{X}_i(t+1)=\hat{X}_i(t)$.    
As for the unreliable channel, we suppose that for user $i$, at each time slot $t$, the probability of having successful transmission is $\rho_i$, and $1-\rho_i$ otherwise. Consequently, the channel realizations are independent and identically distributed
(i.i.d.) over time slots that we denote $c_i(t)$, i.e. $c_i(t)=1$ if the packet is successfully transmitted and $c_i(t)=0$ otherwise.

The next aspect of our model that we tackle is the nature of the process $X_i(t)$.
To that extent, for each user $i$, the information process of interest $X_i(t)$ evolves under Markov chain. For that, we define the probability of remaining at the same state in the next time slot as $p_i$. Similarly, the probability of transitioning to another state is $r_i$. Denoting by $N_i$ the number of possible states of $X_i(t)$, then the following always holds:
\begin{equation}\label{eq:relation_p_r}
p_i+(N_i-1)r_i=1
\end{equation}
In this chapter, We study the case where $p_i \geq r_i$.
\begin{figure}[H]
\centering
\includegraphics[width=0.4\textwidth]{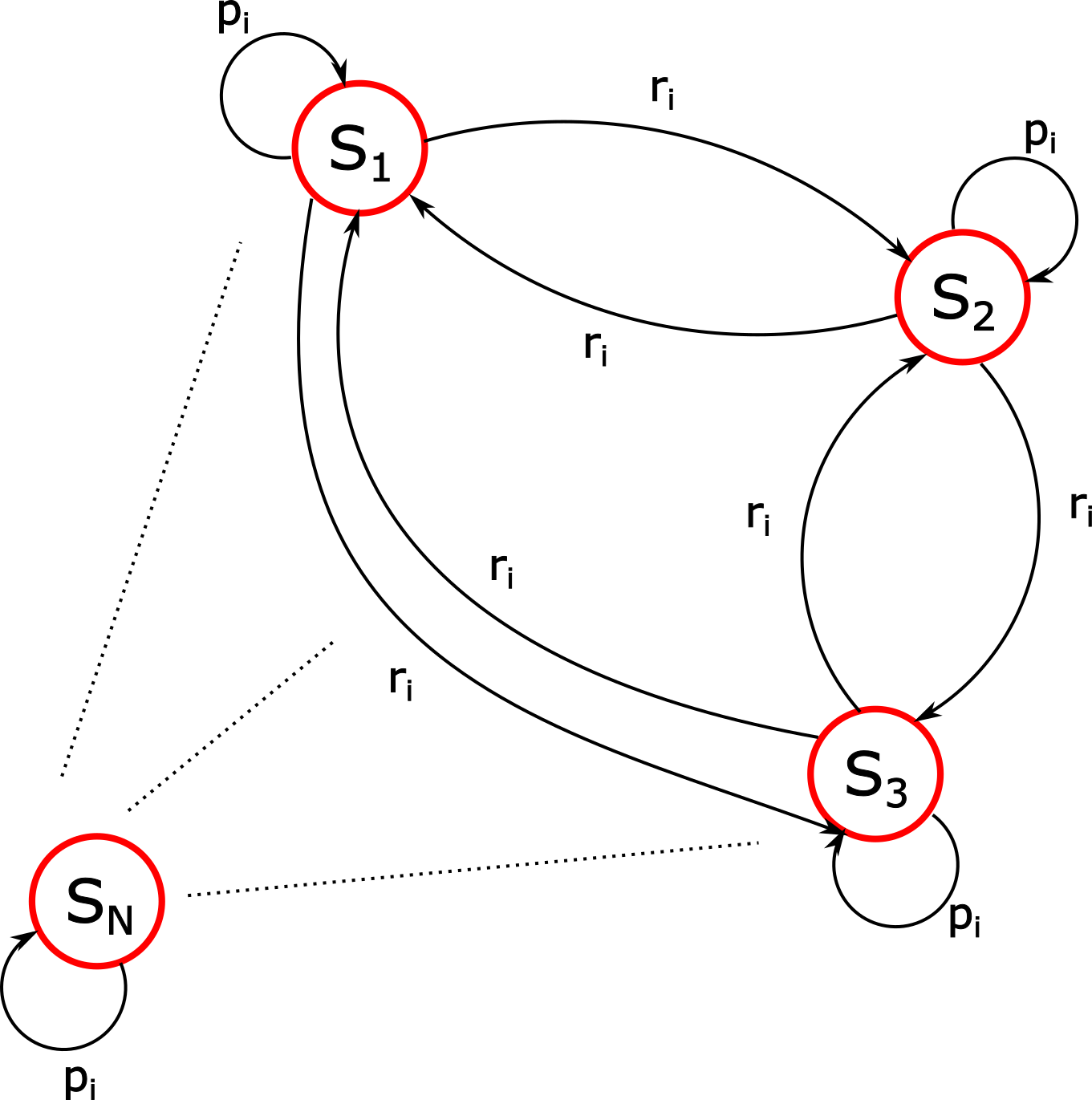}
\caption{Illustration of process $X_i(t)$}
\end{figure}
\subsection{Penalty function dynamics}
In this paper, we study the mean age of incorrect information (MAoII) penalty function and we compare it with the age of information metric (AoI). We see how it is relevant, accurate and more realistic to consider MAoII metric compared to AoI metric in order to have a good performance with regard to the empirical value of the age of information. For that purpose, we start by reintroducing in the next section the age of information to emphasize its shortcomings, then we propose as an alternative metric MAoII. 
\subsubsection{Age of information penalty function}
The standard metric (AoI) that captures the freshness of information for user $i$ is: 
\begin{equation}
\delta_{AoI}(t)=t-g_i(t)
\end{equation}
where $g_i(t)$\footnote{Considering our system model detailed in \ref{subsec:Net_descrip}, $g_i(t)$ refers also to the sampling time of the information of interest contained in the last successfully received packet} is the time-stamp of the last successfully received packet by monitor. This metric captures the lifetime of the last update at the monitor without taking into account the correctness of the information. Thereby, this makes it fall short in some applications. For instance in some scenarios, the age will increase but the information of interest remains at the same state. Nevertheless, to further emphasize the shortcoming of this metric, we provide the Whittle index policy considering this metric which is already derived in \cite{maatouk2019age}. And we give some numerical results that show the shortage of this policy.   

\subsubsection{Mean Age of incorrect information penalty function}\label{subsubsec:MAoII_pf} 
The age of incorrect information has been introduced the first time in \cite{maatouk2019age}. This metric captures the freshness of informative updates. Specifically, if the monitor acquires the information about the process $X_i(t)$, as long as the state of the process $X_i(t)$ remains at the same state in the next time slots, the age of the incorrect information will not increase, since there is no new information unknown by the monitor. In \cite{maatouk2019age}, the authors presume that the scheduler has a perfect knowledge of the process at each time slot and restrict their analysis to a transmitter-receiver pair communication. While in our case, we consider that the monitor which plays the role of the scheduler, knows only the state of the last successively received packet and we extend our analysis to a communication involving several users that can transmit at each time slot. Accordingly, the explicit expression of MAoII metric is:
\begin{equation}
\delta_{MAoII}(t)=\mathbb{E}_{V_i}[(t-V_i(t)]
\end{equation}
where $V_i(t)$ is the last time instant such that $\mathbf{1}_{\{X_i(V_i(t))= \hat{X}_i(g_i(t)+1)\}}=1$.
\begin{remark}
It is worth mentioning that, as it was explained in Section \ref{subsec:Net_descrip}, the reception of the successfully transmitted packet takes place at time slot $g_i(t)+1$. This means that $\hat{X}_i(g_i(t)+1)=X_i(g_i(t))$.
\end{remark}
In order to use this metric effectively in a partially Observable
Markov Decision Process Problem, we need to take into consideration the markovian nature of the process $X_i(t)$. To that extent, we introduce in the next section the notion of the belief that represents the probability that $\hat{X}_i(t)$ is in the correct state.\\
 
\subsection{Metrics evolution}\label{subsec:metrics_evolution}
In this section, we describe mathematically the evolution of each metric depending on the system parameters and the action taken.
We denote by $d_i(t)$ the action prescribed to user $i$ at time slot $t$ and by $a_i$, $b_i$, the age of information and the mean age of incorrect information penalty functions respectively. 
\subsubsection{AoI}\label{subsubsec:AoI_evolution}
Considering our system model, the age of information of user $i$ evolves as follows:
If the user $i$ is scheduled ($d_i(t)=1$), the value of AoI goes to state $1$ if the packet is successively transmitted ($c_i(t)=1$), otherwise, the value of AoI is increased by one ($c_i(t)=0$).
If the user $i$ is not scheduled ($d_i(t)=0$), the value of AoI is increased by one.
Accordingly, the evolution of the age of the user $i$ can be summarized in the following:
\begin{equation}
 a_i(t+1)=\left\{
    \begin{array}{ll}
        1 & if d_i(t)=1, c_i(t)=1 \\
        a_i(t)+1 & else
    \end{array}
\right.
\end{equation}
As for the second metric, to highlight the notion of correctness, the monitor maintains a belief value $\pi_i(t)$ which is defined as the probability that the information state in the monitor, $\hat{X_i}(t)=\hat{X_i}(g_i(t)+1)=X_i(g_i(t))$\footnote{We recall that as long as the monitor has not received any new update from the source at time instant $t$, it maintains the last update successively received at time instant $g_i(t)+1$. In other words, $\hat{X_i}(t)=\hat{X_i}(g_i(t)+1)$} at time $t$ being correct. Explicitly $\pi_i(t)=Pr(\hat{X_i}(t)=X_i(t))$. 
One can show that $\pi_i(t)$ evolves as follows:
\begin{lemma}\label{lem:pi_evolution}
\begin{equation}
 \pi_i(t+1)=\left\{
    \begin{array}{ll}
        p_i & if d_i(t)=1, c_i(t)=1 \\
        \pi_i(t)p_i+r_i(1-\pi_i(t)) & else
    \end{array}
\right.
\end{equation}
\end{lemma}
\begin{IEEEproof}
See appendix \ref{app:lem:pi_evolution}.

\end{IEEEproof}

\subsubsection{MAoII}\label{subsubsec:MAoII_evolution}
According to the expression of MAoII given in section \ref{subsubsec:MAoII_pf}, $(t-V_i(t))$ is a random variable that we denote $A_i(t)$ that satisfies:
\begin{lemma}\label{lem:random_variable}
\begin{align}
A_i(t)=\left\{
    \begin{array}{lll}
        0 & w.p & \pi_i(t)\\
        1 & w.p &  \pi_i(t-1).(1-p_i)\\
        2 & w.p &  \pi_i(t-2).(1-p_i).(1-r_i)\\
        3 & \cdots  &         \cdots\\
        \vdots & &\\
        t-g_i(t)-1& w.p & \pi_i(g_i(t)+1).(1-p_i)\\
                  &  & \  .(1-r_i)^{t-g_i(t)-2} \\
        \\          
        t-g_i(t)& w.p & (1-p_i).(1-r_i)^{t-g_i(t)-1} 
    \end{array}
\right.
\end{align}
\end{lemma}
\begin{IEEEproof}
See appendix \ref{app:lem:random_variable}.
 
\end{IEEEproof}
Therefore, the mean of the age of the incorrect information at slot $t$ equals to the mean of $A_i(t)$, i.e.
\begin{align}
n_i(t)=&\mathbb{E}[A_i(t)] \nonumber \\
=&\sum_{k=0}^{t-g_i(t)-1} k(1-p_i)(1-r_i)^{k-1} \pi_i(t-k) \nonumber \\
&+(t-g_i(t)).(1-p_i).(1-r_i)^{t-g_i(t)-1} \nonumber \\
=&\sum_{k=1}^{t-g_i(t)} (t-g_i(t)-k)(1-p_i)(1-r_i)^{t-g_i(t)-k-1} \pi_i(g_i(t)+k)\nonumber \\
&+ (t-g_i(t)).(1-p_i).(1-r_i)^{t-g_i(t)-1} 
\end{align}
One can establish that for all $t$, using definition of $g_i(t)$, $\pi_i(g_i(t)+1)=p_i$. Hence, according to the evolution of $\pi_i(\cdot)$ in Lemma \ref{lem:pi_evolution}, for all $k \leq t$, $\pi_i(g_i(t)+k)$ depends only on $k$ and $i$. More precisely, we have that for each $k\leq t$, $\pi_i(g_i(t)+k)=\pi_i^k$ where $\pi_i^k$ is a sequence defined by induction as follows:
\begin{equation}
 \pi^k_i=\left\{
    \begin{array}{ll}
        1 & if k=0 \\
        p_i \pi_i^k+r_i(1-\pi_i^k) & if k>0
    \end{array}
\right.
\end{equation}
In light of that fact, we have that:
\begin{equation}
n_i(t)=\sum_{k=0}^{t-g_i(t)} (t-g_i(t)-k)(1-p_i)(1-r_i)^{t-g_i(t)-k-1} \pi_i^k
\end{equation}
We conclude that $n_i(t)$ depends on $t-g_i(t)$ and $i$. Therefore, we let $n_i(t)\overset{\Delta}{=}n_i(t-g_i(t))$.\\

To that extent, at time slot $t$, if the user $i$ is scheduled and the packet is successively transmitted, then $g_i(t+1)=t$. Accordingly, at time slot $t+1$, MAoII equals to $n_i(t+1-g_i(t+1))=n_i(1)$.   
If the user $i$ is not scheduled or if the packet is not successively transmitted, then $g_i(t+1)=g_i(t)$. Therefore, MAoII will transit to $n_i(t+1-g_i(t+1))=n_i(t-g_i(t)+1)$. Based on this and denoting $j(t)$ the index such that $n_i(j(t))$ is the value of MAoII at time slot $t$, MAoII will transit to the value $n_i(j(t)+1)$ at time instant $t+1$. To sum up, the evolution of MAoII can be summarized as follows:     
\begin{equation}
 b_i(t+1)=\left\{
    \begin{array}{ll}
        n_i(1) & if d_i(t)=1, c_i(t)=1 \\
        n_i(j(t)+1) & else
    \end{array}
\right.
\end{equation}
where $b_i(t)=n_i(j(t))$.

\section{Problem formulation}\label{sec:prob_form}
In this section, we consider a given metric denoted by $m$ where $m$ can be either AoI or MAoII. We denote further by $N_u$ the total number of users in the system.   
We let the vector $\boldsymbol m$ at time $t$ be $\boldsymbol{m}(t)=(m_{1}(t),\ldots,m_{N_u}(t))$ where $m_i(t)$ is the penalty function at the central entity of user $i$ with respect to the metric $m$ at time slot $t$. Our aim is to find a scheduling policy that allocates per each time slot, the available channels ($M$ channels) to a given subset of users ($M$ users, $M \leq N_u$) in a such way to minimize the total expected average penalty function of the metric considered. A scheduling policy $\phi$ is defined as a sequence of actions $\phi=(\boldsymbol{d}^{\phi}(0),\boldsymbol{d}^{\phi}(1),\ldots)$ where $\boldsymbol{d}^{\phi}(t)=(d_1^{\phi}(t),d_2^{\phi}(t),\ldots,d_{N_u}^{\phi}(t))$ is a binary vector such that $d_i^{\phi}(t)=1$ if the user $i$ is scheduled at time $t$. 
Denoting by $\Phi$, the set of all causal scheduling policies, then
our scheduling problem can be formulated as follows:
\begin{equation}
\setlength{\belowdisplayskip}{0pt} \setlength{\belowdisplayshortskip}{0pt}
\setlength{\abovedisplayskip}{0pt} \setlength{\abovedisplayshortskip}{0pt} 
\begin{aligned}
& \underset{\phi\in \Phi}{\text{minimize}}
& & \lim_{T\to+\infty} \text{sup}\:\frac{1}{T}\mathbb{E}^{\phi\in \Phi}\Big(\sum_{t=0}^{T-1}\sum_{i=1}^{N_u}m_i^{\phi}(t)|\boldsymbol{m}(0)\Big)\\
& \text{subject to}
& & \sum_{i=1}^{N_u}d_{i}^{\phi}(t)\leq\alpha N_u \quad t=1,2,\ldots
\end{aligned}
\label{eq:original_problem}
\end{equation}
where $\alpha N_u=M$.
The problem in (\ref{eq:original_problem}) falls into Multi-armed bandit problems and especially Restless Bandit framework.
RMAB problems are known to be generally difficult to solve them as they are PSPACE-Hard \cite{papadimitriou1999complexity}. To circumvent this complexity, a well-known heuristic is proposed for these types of problems called Whittle's index policy \cite{weber1990index}. This policy is based on a Lagrangian relaxation, and was shown to have remarkable performance in real-life applications. 
To that extent, we tackle in more depth in the next section the Lagrangian relaxation approach applied to our RBP problem. Then, we provide the theoretical analysis to get the low-complex policy: Whittle index policy that we denote by WIP for the two metrics namely, AoI and MAoII.   

\section{Lagrangian Relaxation and Whittle's Index}\label{sec:lag_relax_whi_ind}
\subsection{Relaxed problem}
The Lagrangian relaxation technique is the key component for defining the Whittle's index scheduling policy. First, it consists of relaxing the constraint on the available resources by letting it be satisfied on average rather than in every time slot. More specifically, we define our Relaxed Problem (\textbf{RP}) as follows:
\begin{equation}\label{eq:relaxed_problem}
\setlength{\belowdisplayskip}{0pt} \setlength{\belowdisplayshortskip}{0pt}
\setlength{\abovedisplayskip}{0pt} \setlength{\abovedisplayshortskip}{0pt} 
\begin{aligned}
& \underset{\phi\in \Phi}{\text{minimize}}
& & \lim_{T\to+\infty} \text{sup}\:\frac{1}{T}\mathbb{E}^{\phi}\Big(\sum_{t=0}^{T-1}\sum_{i=1}^{N_u}m_i^{\phi}(t)|\boldsymbol{m}(0)\Big)\\
& \text{subject to}
& & \lim_{T\to+\infty}\text{sup}\frac{1}{T}\mathbb{E}^{\phi}\Big(\sum_{t=0}^{T-1}\sum_{i=1}^{N_u}d_{i}^{\phi}(t)\Big)\leq\alpha N_u
\end{aligned}
\end{equation}
\color{black}
The Lagrangian function $f(W,\phi)$ of the problem \eqref{eq:relaxed_problem} is defined as:
\begin{equation}
\lim_{T\to+\infty} \text{sup}\:\frac{1}{T}\mathbb{E}^{\phi}\Big(\sum_{t=0}^{T-1}\sum_{i=1}^{N_u}m_i^{\phi}(t)+Wd_{i}^{\phi}(t)|\boldsymbol{m}(0)\Big)-W\alpha N_u
\end{equation}
where $W \geq 0$ can be seen as a penalty for scheduling users. Thus, by following the Lagrangian approach, our next objective is to solve the following problem:

\begin{equation}\label{eq:dual_problem}
\\\\ \underset{\phi\in \Phi}{\text{min}} f(W,\phi)
\end{equation}

As the term $W\alpha N_u$ is independent of $\phi$, it can be eliminated from the analysis. Baring that in mind, we present the steps to obtain the Whittle's index policy:
\begin{enumerate}
\item We focus on the one-dimensional version of the problem in (\ref{eq:dual_problem}). Indeed, it can be shown that the $N_u$-dimensional problem can be decomposed into $N_u$ one-dimensional problems that can be solved independently \cite{kriouile2018asymptotically}. Accordingly, we drop the user's index for ease of notation involving all user's parameters, and we deal with the one-dimensional problem:
\begin{equation}\label{eq:individual_dual_problem}
\setlength{\belowdisplayskip}{0pt} \setlength{\belowdisplayshortskip}{0pt}
\setlength{\abovedisplayskip}{0pt} \setlength{\abovedisplayshortskip}{0pt} 
\begin{aligned}
& \underset{\phi\in \Phi}{\text{min}}
& & \lim_{T\to+\infty} \text{sup}\:\frac{1}{T}\mathbb{E}^{\phi}\Big(\sum_{t=0}^{T-1}m^{\phi}(t)+Wd^{\phi}(t)|m(0)\Big)
\end{aligned}
\end{equation}
\item We give the structural results on the optimal solution of the one-dimensional problem.
\item We establish the indexability property, which ensures the existence of the Whittle's indices.
\item We derive a closed-form expression of the Whittle's index and, thereby, define the proposed scheduling policy (WIP) for the original problem (\ref{eq:original_problem}).
\end{enumerate}
\subsection{Structural results}
The problem in (\ref{eq:individual_dual_problem}) can be viewed as an infinite horizon average cost Markov decision process that is defined as follows:
\begin{itemize}
\item \textbf{States}: The state of the MDP at time $t$ is the penalty function $m(t)$. 
\item \textbf{Actions}: The action at time $t$, denoted by $d(t)$, specify if the user is scheduled (value $1$) or not (value $0$).
\item \textbf{Transitions probabilities}: The transitions probabilities between the different states.
\item \textbf{Cost}: We let the instantaneous cost of the MDP, $C(m(t),d(t))$, be equal to $m(t)+Wd(t)$.
\end{itemize}
The optimal policy $\phi^*$ of the one-dimensional problem \eqref{eq:individual_dual_problem} can be obtained by solving the following Bellman equation for each state $m$:
\begin{align}
\theta &+ V(m) \nonumber \\ 
&=\min_{d\in\{0,1\}}\big\{m+Wd+\sum_{m'\in A^m }\Pr(m\rightarrow m'|d)V(m')\big\} 
\label{eq:bellman_general}
\end{align}
where $\Pr(m\rightarrow m'|d)$ is the transition probability from state $m$ to $m'$ under action $d$, $\theta$ is the optimal value of the problem, $V(m)$ is the differential cost-to-go function and $A^m$ is the set of states of the metric $m$. There exist several numerical algorithms that are developed to solve (\ref{eq:bellman_general}), such as the value iteration algorithm. This later consists first of updating per each iteration the value function $V_t(.)$ following the recurrence relation for each state $m$:
\begin{align}\label{eq:bellman_equation_time_t}
\theta &+ V_{t+1}(m)\\
&=\min_{d\in\{0,1\}}\big\{m+Wd+\sum_{m'\in A^m }\Pr(m\rightarrow m'|d)V_t(m')\big\}  
\end{align}
given that $V_0(.)=0$.
Then concluding for $V(.)$ exploiting the fact that $\underset{t \rightarrow +\infty}{\text{lim}} V_t(m)=V(m)$. The main shortcoming of this algorithm that it requires high memory and computational complexity. To overcome this complexity, rather than computing the value of $V(.)$ for all states, we limit ourself to study the structure of the optimal scheduling policy by exploiting the fact that $\underset{t \rightarrow +\infty}{\text{lim}} V_t(m)=V(m)$. In that way, we show that the optimal solution of Problem \eqref{eq:bellman_general} is a threshold-based policy: 
\begin{definition} 
A threshold policy is a policy $\phi \in \Phi$ for which there exists $n$ such that when the current state $m < n$, the prescribed action is $d^- \in \{0,1\}$, and when $ m \geq n$, the prescribed action is $d^+ \in \{0,1\}$ while baring in mind that $d^- \neq d^+$.
\end{definition}
To that extent, we show that for both metrics considered in our paper, namely AoI and MAoII, the optimal policy of \eqref{eq:bellman_general} is a threshold based policy. 
To that end, we specify first the states space $A^m$ for both metrics, then we provide the expression of the corresponding Bellman equation \eqref{eq:bellman_general}. After that, we establish our desired result.
\subsubsection{AoI}
According to Section \ref{subsubsec:AoI_evolution}, $a(t)$ evolves in the state space:
\begin{equation}
A^a=\{a^{j}: j>0, a^{j}=j\}
\end{equation}
The corresponding Bellman equation is:
\begin{align}
\theta_a + V(a^{j})=\min\big\{&a^{j}+V(a^{j+1}); \nonumber \\
&a^{j}+W+\rho  V(a^{1})+(1-\rho)V(a^{j+1})\big\} 
\end{align}
The analysis have been already done in \cite{maatouk2020optimality} regarding this metric. Effectively, in \cite{maatouk2020optimality}, the authors demonstrate that the structure of the optimal policy of Problem \eqref{eq:individual_dual_problem} is a threshold based policy. They prove further that this policy is increasing with the age. i.e.:
\begin{proposition}
When $m=a$, the optimal solution of the problem in (\ref{eq:individual_dual_problem}) is an increasing threshold policy. Explicitly, there exists $n$ such that when the current state $a^{j} < a^{n}$, the prescribed action is a passive action, and when $ a^{j} \geq a^{n}$, the prescribed action is an active action.
\end{proposition}
One can see the detailed proof in \cite{maatouk2020optimality}.
\subsubsection{MAoII}
According to Section \ref{subsubsec:MAoII_evolution}, $b(t)$ evolves in the state space:
\begin{equation}
A^b=\{b^{j}: j>0, b^{j}=\sum_{k=0}^{j} k(1-p)(1-r)^{k-1} \pi^{j-k}\}
\end{equation}
Therefore, the expression of Bellman equation at state $b^{j}$ 
\begin{align}
\theta_b + V(b^{j})=\min\big\{&b^{j}+V(b^{j+1});\nonumber \\
&b^{j}+W+\rho  V(b^{1})+(1-\rho)V(b^{j+1})\big\} 
\end{align}

\begin{theorem}\label{theo:threshold_policy_2}
When $m=b$, the optimal solution of the problem in (\ref{eq:individual_dual_problem}) is an increasing threshold policy. Explicitly, there exists $n$ such that when the current state $b^{j} < b^{n}$, the prescribed action is a passive action, and when $ b^{j} \geq b^{n}$, the prescribed action is an active action.
\end{theorem}
\begin{IEEEproof}
The proof can be found in Appendix \ref{app:theo:threshold_policy_2}.
\end{IEEEproof}
\subsection{Indexability and Whittle's index expressions}
In order to establish the indexability of the problem and find the Whittle's index expressions, we provide the steady-state form of the problem in (\ref{eq:individual_dual_problem}) under a threshold policy $n$. Explicitly:
\begin{equation}
\begin{aligned}
& \underset{n\in \mathbb{N}^*}{\text{minimize}} 
& & \overline{m^{n}}+W\overline{d^n}
\end{aligned}
\label{thresholdobjective}
\end{equation}
where $\overline{m^{n}}$ is the average value of the penalty function with the respect to the metric $m$, and $\overline{d^n}$ is the average active time under threshold policy $n$. Specifically: 
\begin{align}
\overline{m^{n}}&=\lim_{T\to+\infty} \text{sup}\:\frac{1}{T}\mathbb{E}^{n}\Big(\sum_{t=0}^{T-1}m(t)|m(0),tp(n)\Big)\label{eq:average_age}\\
\overline{d^n}&=\lim_{T\to+\infty} \text{sup}\:\frac{1}{T}\mathbb{E}^{n}\Big(\sum_{t=0}^{T-1}d(t)|m(0),tp(n)\Big)\label{eq:average_active_time}
\end{align}
where $tp(n)$ denotes the threshold policy $n$.
With the intention of computing $\overline{m^{n}}$ and  $\overline{d^n}$, we derive the stationary distribution of the Discrete Time Markov Chain, DTMC that represents the evolution of MAoII under threshold policy $n$. One can show that the steady state distribution in question is the same for both metrics, AoI and MAoII. Specifically: 
\begin{proposition}\label{prop:stationary_distribution}
For $m=a,b$, for a given threshold $n$, the DTMC admits $u^n(m^{j})$ as its stationary distribution:
\begin{equation}
 u^n(m^{j})=\left\{
    \begin{array}{ll}
        \frac{\rho  }{n\rho  +1-\rho  } & \text{if} \ 1 \leq j \leq n  \\
        (1-\rho  )^{j-n} \frac{\rho  }{n\rho  +1-\rho  } & \text{if} \ j \geq n\\
    \end{array}
\right.
\end{equation}
\label{stationarydistribution}
\end{proposition}
\vspace{-10pt}
\begin{IEEEproof}
The proof can be found in Appendix \ref{app:prop:stationary_distribution}.
\end{IEEEproof}

By exploiting the above results, we can now proceed with finding a closed-form of the average cost of any threshold policy.

\begin{proposition}
For a given threshold $n$, the average cost of the threshold policy is $\overline{m^{n}}$:
\begin{itemize}
\item $m=a$:
\begin{align}
\overline{a^{n}}=&\frac{[(n-1)^2+(n-1)]\rho  ^2+2\rho  (n-1)}{2\rho  ((n-1)\rho  +1)}\nonumber\\&+\frac{2}{2\rho  ((n-1)\rho+1)}
\label{costgamma}
\end{align}
\item $m=b$:
\begin{align}
\overline{b^{n}}=&\frac{\rho  }{n\rho  +1-\rho  }[\frac{n(N-1)}{Nr}-\frac{(1-Nr)^{n+2}}{(Nr)^2}\nonumber\\
&+\frac{(1-r)^{n+2}}{r^2}+\frac{(1-\rho  )(1-Nr)^{n+2}}{Nr(1-(1-\rho  )(1-r))}-\nonumber\\
&\frac{(1-\rho  )(1-r)^{n+2}}{r(1-(1-\rho  )(1-r))}+C]
\end{align}
\end{itemize}
where $C=\frac{(1-Nr)^2}{(Nr)^2}-\frac{(1-r)^2}{r^2}+\frac{(N-1)(1-\rho )}{Nr\rho}$.
\end{proposition}
\begin{IEEEproof}
By leveraging the results of Proposition \ref{prop:stationary_distribution} and using the expression of $m^{j}$ for $j>0$, by definition of $\overline{m^{n}}$ given in \eqref{eq:average_age}, we get after algebraic manipulations the desired results.
\end{IEEEproof}

\begin{proposition}
For any given threshold $n$, the active average time is $\overline{d^n}$:
\begin{align}
\overline{d^n}=&\frac{1}{n\rho  +1-\rho  }
\end{align}
\end{proposition}
\begin{IEEEproof}
Likewise, exploiting the results in Proposition \ref{prop:stationary_distribution} and according to the expression \eqref{eq:average_active_time}, we obtain the desired results.
\end{IEEEproof}

To ensure the existence of the Whittle's indices, we need first to establish the indexability property for all users.
To that end, we first formalize the indexability and the Whittle's index in the following definitions.
We note that in the sequel, we precise the indices of users to differentiate between them.

\begin{definition}
Considering Problem~\eqref{eq:individual_dual_problem} for a given $W$ and a given user $i$, we define $D_i^m(W)$ as the set of states in which the optimal action (with respect to the optimal solution of Problem~\eqref{eq:individual_dual_problem} considering the metric $m$) is the passive one. In other words, $m_i^{n} \in D_i^m(W)$ if and only if the optimal action at state $m_i^{n}$ is the passive one.
\end{definition}
$D_i^m(W)$ is well defined for both metrics as the optimal solution of Problem~\eqref{eq:individual_dual_problem} is a stationary policy, more precisely, a threshold based policy.
\begin{definition}\label{def:Whitt_index}
A class is indexable if the set of states in which the passive action is the optimal action increases with $W$, that is, $W' < W \Rightarrow D_i^m(W') \subseteq D_i^m(W)$.
When the class is indexable, the Whittle's index in state $m_i^{n}$ is defined as: 
\begin{equation}
W(m_i^{n})=\min \{W |m_i^{n} \in D_i^m(W)\}
\end{equation} 
\end{definition}
\begin{proposition}
For each user $i$, the one-dimensional problem is indexable for both metrics.
\end{proposition}
\begin{IEEEproof}
The proof rests on the decrease of $\overline{d_i^n}$ with $n$. One can see \cite{maatouk2020optimality} for a detailed proof.
\end{IEEEproof}
As the indexability property has been established in the above proposition, we can now assert the existence of the Whittle's index.
\begin{theorem}\label{theo:Whittle_index_expressions}
For any user $i$ and state $m_i^{n}$, the Whittle's index is:
\begin{itemize}
\item $m=a$
\begin{equation}
W_i(a_i^{n})=\frac{n(n-1)\rho_i}{2}+n
\end{equation}
\item $m=b$
\begin{align}
W_i(b_i^{n})=&\frac{(1-r_i)^2 \rho_i}{r_i^2}-\frac{(1-N_ir_i)^2 \rho_i}{(N_ir_i)^2} \nonumber \\ 
&+(1-N_ir_i)^{n+2}(n \rho_i+1+\frac{\rho_i(1-N_i r_i)}{N_i r_i}) \nonumber \\
& \ \ \ \times [\frac{1-(1-\rho_i)(1+(N_i-1)r_i)}{N_i r_i (1-(1-\rho_i)(1-r_i))} ]\nonumber \\
&-(1-r_i)^{n+2}(n \rho_i+1+\frac{\rho_i(1-r_i)}{r_i}) \nonumber \\
& \ \  \ \times [\frac{\rho_i}{ r_i (1-(1-\rho_i)(1-r_i))}]
\end{align}
\end{itemize}
\end{theorem}
\begin{IEEEproof}
The proof can be found in Appendix \ref{app:theo:Whittle_index_expressions}.
\end{IEEEproof}
Based on the above proposition, we provide in the following the Whittle's index scheduling policy for the original problem \eqref{eq:original_problem}.
\begin{algorithm}
\caption{Whittle's index scheduling policy}\label{euclid}
\begin{algorithmic}[1]
\State At each time slot $t$, compute the Whittle's index of all users in the system using the expressions given in Proposition \ref{theo:Whittle_index_expressions}.
\State Allocating the $M$ channels to the $M$ users having the highest Whittle's index values at time $t$.
\end{algorithmic}
\end{algorithm}\\

\section{Numerical Results}\label{sec:num_reslt}
Our goal in this section is to compare the average empirical age of incorrect information under the developed Whittle index policy WIP-MAoII to the baseline policy, denoted by WIP-AoI, that considers the standard AoI metric.
More precisely, we plot  $C^{\phi,N_u}=\frac{1}{N_u}\lim_{T\to+\infty} \text{sup}\:\frac{1}{T}\mathbb{E}^{\phi}\Big(\sum_{t=0}^{T-1}\sum_{i=1}^{N_u}m_i^{emp,\phi}(t)|\boldsymbol{m}^{emp}(0),\phi\Big)$ for $\phi$ equals to WIP-MAoII and WIP-AoI, in function of $N_u$, where $m^{emp,\phi}_i(\cdot)$ evolves as follows:
\begin{itemize}
\item If $m_i^{emp,\phi}(t)=0$, then $\hat{X}_i(t)=X_i(t)$. Therefore:
\begin{equation}
 m_i^{emp,\phi}(t+1)=\left\{
    \begin{array}{lll}
        0 & w.p & p_i \\
        1 & w.p & 1-p_i \\ 
    \end{array}
\right.
\end{equation}
\item If $m_i^{emp,\phi}(t) \neq 0$, then $\hat{X}(t) \neq X(t)$. Therefore:
\begin{itemize}
\item If $\phi_i(t)=1$,
\begin{equation}
 m_i^{emp,\phi}(t+1)=\left\{
    \begin{array}{lll}
        0 & w.p & \rho_i p_i \\
        1 & w.p & \rho_i (1-p_i) \\
        m_i^{emp,\phi}(t)+1 & w.p & (1-\rho_i) \\
                            &      & \times (1-r_i) \\
        0 & w.p & (1-\rho_i)r_i \\  
    \end{array}
\right.
\end{equation}
\item If $\phi_i(t)=0$,
\begin{equation}
 m_i^{emp,\phi}(t+1)=\left\{
    \begin{array}{lll}
        m_i^{emp,\phi}(t)+1 & w.p & (1-r_i) \\
        0 & w.p & r_i \\  
    \end{array}
\right.
\end{equation}
\end{itemize}
\end{itemize}
We consider two scenarios of the network settings:
\begin{enumerate}
\item For the first scenario, we consider two classes with the respective parameters:
\begin{itemize}
\item Class 1: $\rho_1=0.7$, $N_1=8$, $r_1=0.1$\footnote{The value of $p_i$ can be directly deduced from Equation \ref{eq:relation_p_r}}.
\item Class 2: $\rho_2=0.5$, $N_2=2$, $r_2=0.4$. 
\end{itemize} 
\item Regarding the second scenario, to shed light on the importance of taking into account the source parameters namely, $p_i$, $r_i$ and $N_i$, in the derivation of Whittle's indices, we consider that the two classes share the same channel statics, specifically $\rho_1=\rho_2$, while they don't have the same source parameters. To that extent, we consider the following case:
\begin{itemize}
\item Class 1:  $\rho_1=0.4$, $N_1=10$, $r_1=0.05$
\item Class 2: $\rho_2=0.4$, $N_2=3$, $r_2=0.3$
\end{itemize} 
\end{enumerate}

\begin{figure}[H] \label{fig:comp_aoi_maoii_diff}
\centering
\includegraphics[scale=0.6]{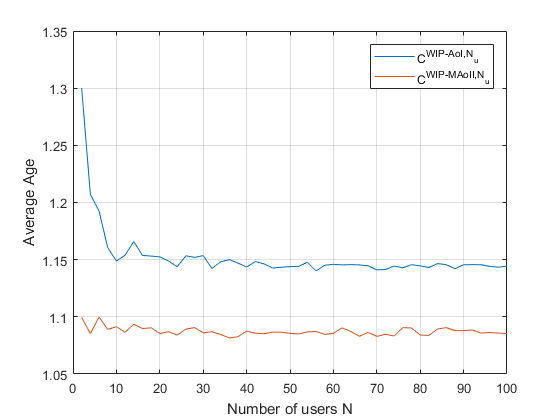}
\caption{Comparison between WIP-MAoII and WIP-AoI in terms of the empirical average age: different channel statics}
\end{figure}
\begin{figure}[H]\label{fig:comp_aoi_maoii_same}
\centering
\includegraphics[scale=0.6]{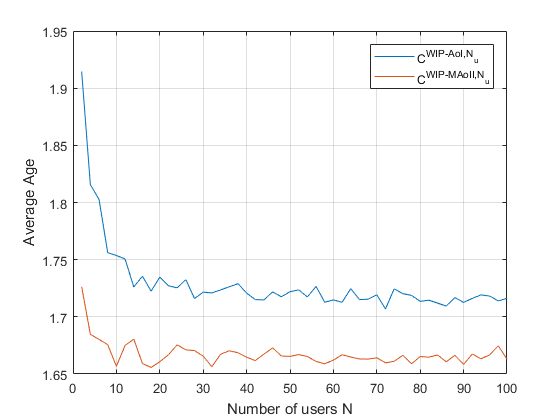}
\caption{Comparison between WIP-MAoII and WIP-AoI in terms of the empirical average age: same channel statics}
\end{figure}

One can observe that effectively WIP-MAoII gives us better performance than WIP-AoI in terms of minimizing the average empirical age of incorrect information. As consequence, our derivation of Whittle's indices in the Markovian source framework turns out to be relevant in terms of tracking the real state of remote sources.

\section{Conclusion}\label{sec:concl}
In this paper, we considered the problem of remote monitoring of multiple sources where a central entity selects at each time a subset of the sources to send their updates, in such a way to minimize the MAoII metrics. Since the scheduler is unaware of the current state of the source, we introduce a belief state at the monitor in order to predict the evolution of the states of the sources and to derive an estimation of the MAoII. We then developed an efficient scheduling policy based on Whittle's index framework. Finally, we have provided numerical results that highlight the performance of our policy.

\bibliographystyle{IEEEtran} 
\bibliography{bibliography}

\begin{thebibliography}{10}
\providecommand{\url}[1]{#1}
\csname url@samestyle\endcsname
\providecommand{\newblock}{\relax}
\providecommand{\bibinfo}[2]{#2}
\providecommand{\BIBentrySTDinterwordspacing}{\spaceskip=0pt\relax}
\providecommand{\BIBentryALTinterwordstretchfactor}{4}
\providecommand{\BIBentryALTinterwordspacing}{\spaceskip=\fontdimen2\font plus
\BIBentryALTinterwordstretchfactor\fontdimen3\font minus
  \fontdimen4\font\relax}
\providecommand{\BIBforeignlanguage}[2]{{%
\expandafter\ifx\csname l@#1\endcsname\relax
\typeout{** WARNING: IEEEtran.bst: No hyphenation pattern has been}%
\typeout{** loaded for the language `#1'. Using the pattern for}%
\typeout{** the default language instead.}%
\else
\language=\csname l@#1\endcsname
\fi
#2}}
\providecommand{\BIBdecl}{\relax}
\BIBdecl

\bibitem{kaul2012real}
S.~Kaul, R.~Yates, and M.~Gruteser, ``Real-time status: How often should one
  update?'' in \emph{2012 Proceedings IEEE INFOCOM}.\hskip 1em plus 0.5em minus
  0.4em\relax IEEE, 2012, pp. 2731--2735.

\bibitem{maatouk2020optimality}
A.~Maatouk, S.~Kriouile, M.~Assaad, and A.~Ephremides, ``On the optimality of
  the whittle's index policy for minimizing the age of information,''
  \emph{arXiv preprint arXiv:2001.03096}, 2020.

\bibitem{hsu2019scheduling}
Y.-P. Hsu, E.~Modiano, and L.~Duan, ``Scheduling algorithms for minimizing age
  of information in wireless broadcast networks with random arrivals,''
  \emph{IEEE Transactions on Mobile Computing}, 2019.

\bibitem{kadota2018scheduling}
I.~Kadota, A.~Sinha, E.~Uysal-Biyikoglu, R.~Singh, and E.~Modiano, ``Scheduling
  policies for minimizing age of information in broadcast wireless networks,''
  \emph{IEEE/ACM Transactions on Networking}, vol.~26, no.~6, pp. 2637--2650,
  2018.

\bibitem{jiang2019unified}
Z.~Jiang, S.~Zhou, Z.~Niu, and C.~Yu, ``A unified sampling and scheduling
  approach for status update in multiaccess wireless networks,'' in \emph{IEEE
  INFOCOM 2019-IEEE Conference on Computer Communications}.\hskip 1em plus
  0.5em minus 0.4em\relax IEEE, 2019, pp. 208--216.

\bibitem{sun2019sampling}
Y.~Sun, Y.~Polyanskiy, and E.~Uysal, ``Sampling of the wiener process for
  remote estimation over a channel with random delay,'' \emph{IEEE Transactions
  on Information Theory}, vol.~66, no.~2, pp. 1118--1135, 2019.

\bibitem{kam2018towards}
C.~Kam, S.~Kompella, G.~D. Nguyen, J.~E. Wieselthier, and A.~Ephremides,
  ``Towards an effective age of information: Remote estimation of a markov
  source,'' in \emph{IEEE INFOCOM 2018-IEEE Conference on Computer
  Communications Workshops (INFOCOM WKSHPS)}.\hskip 1em plus 0.5em minus
  0.4em\relax IEEE, 2018, pp. 367--372.

\bibitem{maatouk2019age}
A.~Maatouk, S.~Kriouile, M.~Assaad, and A.~Ephremides, ``The age of incorrect
  information: A new performance metric for status updates,'' \emph{arXiv
  preprint arXiv:1907.06604}, 2019.

\bibitem{maatouk2020age}
A.~Maatouk, M.~Assaad, and A.~Ephremides, ``The age of incorrect information:
  an enabler of semantics-empowered communication,'' \emph{arXiv preprint
  arXiv:2012.13214}, 2020.

\bibitem{kam2020age}
C.~Kam, S.~Kompella, and A.~Ephremides, ``Age of incorrect information for
  remote estimation of a binary markov source,'' in \emph{IEEE INFOCOM
  2020-IEEE Conference on Computer Communications Workshops (INFOCOM
  WKSHPS)}.\hskip 1em plus 0.5em minus 0.4em\relax IEEE, 2020, pp. 1--6.

\bibitem{papadimitriou1999complexity}
C.~H. Papadimitriou and J.~N. Tsitsiklis, ``The complexity of optimal queuing
  network control,'' \emph{Mathematics of Operations Research}, vol.~24, no.~2,
  pp. 293--305, 1999.

\bibitem{weber1990index}
R.~R. Weber and G.~Weiss, ``On an index policy for restless bandits,''
  \emph{Journal of Applied Probability}, vol.~27, no.~3, pp. 637--648, 1990.

\bibitem{kriouile2018asymptotically}
S.~Kriouile, M.~Larranaga, and M.~Assaad, ``Asymptotically optimal delay-aware
  scheduling in wireless networks,” arxiv e-prints, p,'' \emph{arXiv preprint
  arXiv:1807.00352}, 2018.

\end{thebibliography}

\begin{appendices}
\section{Proof of Lemma \ref{lem:pi_evolution}}\label{app:lem:pi_evolution}
We start by proving the first statement: if $d_i(t)=1$ and $c_i(t)=1$, then the central entity acquires the timeless information about the process $X_i(t)$, it knows the effective state of the process that we denote by $S$ at time $t+1$. Specifically, $\hat{X_i}(t+1)=X_i(t)=S$. Therefore, the probability that $X_i(t+1)=\hat{X_i}(t+1)=S$ knowing the fact that $X_i(t)=S$ is exactly the probability of remaining at the same state. Consequently, $\pi_i(t+1)=p_i$.\\
If $d_i(t)=1$ and $c_i(t)=0$, or $d_i(t)=0$, then the information state in the monitor side does not change ($\hat{X_i}(t+1)=\hat{X_i}(t)$). Thus, the probability of the event $X_i(t+1)=\hat{X_i}(t+1)$:
\begin{align}
Pr&(X_i(t+1)=\hat{X_i}(t+1)) \nonumber \\
=&Pr(X_i(t+1)=\hat{X_i}(t+1)||X_i(t)=\hat{X_i}(t)) \nonumber \\
& \ \ \ \times Pr(X_i(t)=\hat{X_i}(t)) \nonumber \\
&+Pr(X_i(t+1)=\hat{X_i}(t+1)||X_i(t)\neq\hat{X_i}(t)) \nonumber \\
&\ \  \ \times Pr(X_i(t)\neq\hat{X_i}(t)) \nonumber \\
=&Pr(X_i(t+1)=\hat{X_i}(t)||X_i(t)=\hat{X_i}(t)) \nonumber \\
&\ \ \ \times Pr(X_i(t)=\hat{X_i}(t)) \nonumber \\
&+Pr(X_i(t+1)=\hat{X_i}(t)||X_i(t)\neq\hat{X_i}(t)) \nonumber \\
&\ \ \ \times Pr(X_i(t)\neq\hat{X_i}(t)) \nonumber \\
=&Pr(X_i(t+1)=\hat{X_i}(t)||X_i(t)=\hat{X_i}(t))\pi_i(t) \nonumber \\
&+Pr(X_i(t+1)=\hat{X_i}(t)||X_i(t)\neq\hat{X_i}(t))(1-\pi_i(t)) 
\end{align}
$Pr(X_i(t+1)=\hat{X_i}(t)||X_i(t)=\hat{X_i}(t))$ is the probability of remaining at the same state in the next time slot that equals to $p_i$. Decomposing $Pr(X_i(t+1)=\hat{X_i}(t)||X_i(t)\neq\hat{X_i}(t))$:
\begin{align}
Pr&(X_i(t+1)=\hat{X_i}(t)||X_i(t)\neq\hat{X_i}(t)) \nonumber \\
=&\sum_{S \neq \hat{X_i}(t)} Pr(X_i(t+1)=\hat{X_i}(t)||X_i(t)\neq\hat{X_i}(t), X_i(t)=S) \nonumber \\
 & \hspace{1.5 cm} \times Pr(X_i(t)=S||X_i(t)\neq\hat{X_i}(t))
\end{align}
$Pr(X_i(t+1)=\hat{X_i}(t)||X_i(t)\neq\hat{X_i}(t), X_i(t)=S)$ is the probability of transitioning from $S$ to $\hat{X_i}(t) \neq S$ that equals to $r_i$. Hence:
\begin{align}
Pr(X_i(t+1)&=\hat{X_i}(t)||X_i(t)\neq\hat{X_i}(t)) \nonumber \\
&=r_i\sum_{S \neq \hat{X_i}(t)}Pr(X_i(t)=S||X_i(t)\neq\hat{X_i}(t)) \nonumber \\
&=r_i
\end{align}
That is, combining the two results:
\begin{equation}
Pr(X_i(t+1)=\hat{X_i}(t+1))=\pi_i(t) p_i+(1-\pi_i(t))r_i
\end{equation}
Therefore, the proof is complete.

\section{Proof of Lemma \ref{lem:random_variable}}\label{app:lem:random_variable}
We have $A_i(t)=(t-V_i(t))$. As $g_i(t)$ is known by the monitor, then it is a fixed constant. Whereas $V_i(t)$ which represents the last time instant such that $\mathbf{1}_{\{X_i(V_i(t))= \hat{X_i}(g_i(t)+1)\}}=1$ is unknown by the monitor. Accordingly, it is viewed as a random variable by the monitor. By definition of $g_i(t)$, $\mathbf{1}_{\{X_i(g_i(t))= \hat{X_i}(g_i(t)+1)\}}=1$, then $V_i(t)$ takes values in $[g_i(t),t]$. To that extent, we distinguish between two cases:
\begin{enumerate}
    \item The event $\{V_i(t)=g_i(t)\}$ implies that:
    \begin{itemize}
        \item $X_i(g_i(t))= \hat{X_i}(g_i(t)+1)$
        \item For all $j \in [g_i(t)+1,t]$, $X_i(j)\neq \hat{X_i}(g_i(t)+1)$
    \end{itemize}
    By definition of $g_i(t)$, the probability of $\{X_i(g_i(t))= \hat{X_i}(g_i(t)+1)\}$ is $1$.\\
    The probability of  $ \{\forall j \in [g_i(t)+1,t]$, $X_i(j)\neq \hat{X_i}(g_i(t)+1)||X_i(g_i(t))= \hat{X_i}(g_i(t)+1)\}$ is $(1-r_i)^{t-g_i(t)-1}(1-p_i)$.\\
    Accordingly, the probability of the event $\{V_i(t)=g_i(t)\}$ is $(1-p_i)(1-r_i)^{t-g_i(t)-1}$.\\
    \item For $k \in [g_i(t)+1,t]$, the event $\{V_i(t)=k\}$ implies that:
    \begin{itemize}
    \item At time $k$, $X_i(k)= \hat{X_i}(g_i(t)+1)$.
    \item For all $j \in [k+1,t]$, $X_i(j)\neq \hat{X_i}(g_i(t)+1)$.
    \end{itemize}
The probability of $\{X_i(k)= \hat{X_i}(g_i(t)+1)=\hat{X_i}(k)\}$ is $\pi_i(k)$.\\
The probability of  $ \{\forall j \in [k+1,t]$, $X_i(j)\neq \hat{X_i}(g_i(t)+1)||X_i(k)= \hat{X_i}(g_i(t)+1)\}$ is $(1-r_i)^{t-k-1}(1-p_i)$.\\
Accordingly, the probability of the event $\{V_i(t)=k\}$ is $\pi_i(k)(1-p_i)(1-r_i)^{t-k-1}$.\\
\end{enumerate}

Given that $A_i(t)=t-V_i(t)$, then $A_i(t)=k$ implies $V_i(t)=t-k$. That is, the probability of the event $\{A_i(t)=k\}$ is:
\begin{itemize}
    \item If $k=t-g_i(t)$:\\
    \begin{equation}
    (1-r_i)^{t-g_i(t)-1}(1-p_i)
    \end{equation}
    \item If $k \in [0,t-g_i(t)[$:\\
    \begin{equation}    
    \pi_i(t-k)(1-r_i)^{k-1}(1-p_i)
    \end{equation} 
\end{itemize}
This concludes the proof.

\section{Proof of theorem \ref{theo:threshold_policy_2}}\label{app:theo:threshold_policy_2}

We provide first an useful lemma. 

\begin{lemma}\label{lem:MAoII_increasing}
$b^{j}$ is increasing with $j$
\end{lemma} 

\begin{IEEEproof}
The explicit expression of $b^{j}$ is:
\begin{align}
b^{j}=&\frac{N-1}{1+r-p}(1-(j+1)(1-r)^{j}+j(1-r)^{j+1}] \nonumber \\
+&\frac{1}{1+r-p}[(p-r)^{j+1}-(j+1)(1-r)^{j}(p-r) \nonumber \\
&+j(1-r)^{j+1}]
\end{align}
Therefore, after some computations and mathematical analysis, we obtain:
\begin{equation}
b^{j+1}-b^{j}=Nr[(1-r)^{j+1}-(p-r)^{j+1}]
\end{equation}
Given that $0 \leq p-r \leq 1-r$, then $(p-r)^{j+1} \leq (1-r)^{j+1}$. Therefore, $(1-r)^{j+1}-(p-r)^{j+1} \geq 0$.  
Hence, $b^{j}$ is increasing with $j$.
\end{IEEEproof}

Based on this lemma, we prove the following lemma.
\begin{lemma}\label{lem:V_increasing}
$V(.)$ is increasing with $b^j$.
\end{lemma}
\begin{IEEEproof}
We prove the present lemma by induction using the Value iteration equation \eqref{eq:bellman_equation_time_t}. In fact, we show that $V_t(\cdot)$ is increasing and we conclude for $V(\cdot)$.\\
As $V_0(.)=0$, then the property holds for $t=0$.
If $V_t(.)$ is increasing with $b$, we show that for $b^{j} \leq b^{i}$, $V_{t+1}^0(b^{j}) \leq V_{t+1}^0(b^{i})$ and $V_{t+1}^1(b^{j}) \leq V_{t+1}^1(b^{i})$ where for each $k \in \mathbf{N}^*$:
\begin{align}
V_{t+1}^0(b^{k})&=b^{k}+V_t(b^{k+1}) \\
V_{t+1}^1(b^{k})&=b^{k}+W+\rho V_t(b^{1})+(1-\rho)V_t(b^{k+1}) 
\end{align}
We have that:
\begin{equation}
V_{t+1}^0(b^{j}) - V_{t+1}^0(b^{i})=b^{j}-b^{i}+(V_{t}(b^{j+1}) - V_{t}(b^{i+1}))
\end{equation}
According to Lemma \ref{lem:MAoII_increasing}, given that $b^{j} \leq b^{i}$, then $j \leq i$. That means $b^{j+1} \leq b^{i+1}$. Therefore, since $V_t(.)$ is increasing with $b^j$, we have that:
$$V_{t+1}^0(b^{j}) - V_{t+1}^0(b^{i}) \leq 0$$
As consequence, $V_{t+1}^0(\cdot)$ is increasing with $b^{j}$.\\
In the same way, we have:
$$V_{t+1}^1(b^{j}) - V_{t+1}^1(b^{i})=b^{j}-b^{i}+(1-\rho)(V_{t}(b^{j+1}) - V_{t}(b^{i+1}))$$
Hence: 
\begin{equation}
V_{t+1}^1(b^{j}) - V_{t+1}^1(b^{i}) \leq 0
\end{equation}
As consequence, $V_{t+1}^1(\cdot)$ is increasing with $b^{j}$.\\
Since $V_{t+1}(.)=\min\{V^0_{t+1}(\cdot),V^1_{t+1}(\cdot)\}$, then $V_{t+1}(.)$ is increasing with $b^j$. Accordingly, we demonstrate by induction that $V_t(.)$ is increasing for all $t$. Knowing that $\underset{t \rightarrow +\infty}{\text{lim}} V_t(b^j)=V(b^j)$, $V(.)$ must be also increasing with $b^j$.
\end{IEEEproof}

We define:
\begin{equation}
\Delta V(b^{j})=V^1(b^j)-V^0(b^j)
\end{equation}
where $\underset{t \rightarrow +\infty}{\text{lim}} V_t^0(b^j)=V^0(b^j)$ and $\underset{t \rightarrow +\infty}{\text{lim}} V_t^1(b^j)=V^1(b^j)$.\\
Subsequently, $\Delta V(b^{j})$ equals to:
\begin{equation}
\Delta V(b^{j})=\rho [\frac{W}{\rho }+V(b^1)-V(b^{j+1})]
\end{equation}
According to Lemma \ref{lem:V_increasing}, $V(.)$ is increasing with $b^{j+1}$. Therefore, $\Delta V(b^{j})$ is decreasing with $b^j$. Hence, there exists $b^{n}$ such that for all $b^j \leq b^{n}$, $\Delta V(b^{j})\geq 0$, and for all $b^j > b^{n}$, $\Delta V(b^{j}) <0$. Given that the optimal action for state $b^j$ is the one that minimizes $\min\{V^0(\cdot),V^1(\cdot)\}$, then for all $b^j \leq b^{n}$, the optimal decision is to stay idle since $\min\{V^0(b^j),V^1(b^j)\}=V^0(b^j)$, and for all $b^j > b^{n}$, the optimal decision is to transmit since $\min\{V^0(b^j),V^1(b^j)\}=V^1(b^j)$. Specifically, as $b^{j}$ is increasing with $j$, there exists $n$ such that for all $j\leq n$, the optimal action is passive action, and for all $j> n$, the optimal action is the active one. 

\section{Proof of Proposition \ref{prop:stationary_distribution}}\label{app:prop:stationary_distribution}
In order to demonstrate this proposition, we need to resolve the full balance equation under threshold policy $n$ at each state $m^{j}$ for $m=a$ and $m=b$:
\begin{equation}
u^n(m^{j})=\sum_{i=1}^{+\infty} pt^n(i \rightarrow j)u^n(m^{i})
\end{equation}
where $pt^n(i \rightarrow j)$ denotes the transitioning probability from the state $m^{i}$ to state $m^{j}$ under threshold policy $n$. 
After some computations, we obtain the desired result which is valid for both $m=a$ and $m=b$.  

\section{Proof of Theorem \ref{theo:Whittle_index_expressions}}\label{app:theo:Whittle_index_expressions}
For $m=a$, since the analysis are same as in \cite{maatouk2020optimality}, we skip this case for sake of space.
When $m=b$, using Definition \ref{def:Whitt_index} to find the Whittle’s index expressions can be tricky and difficult. To circumvent this, we first define the sequence $W_i(b_i^{n})$
as the intersection points between $\overline{b_i^n}+W\overline{d_i^n}$ and $\overline{b_i^{n+1}}+W\overline{d_i^{n+1}}$. Explicitly:
\begin{equation}
W_i(b_i^{n})=\frac{\overline{b_i^{n+1}}-\overline{b_i^{n}}}{\overline{d_i^n}-\overline{d_i^{n+1}}}
\end{equation}
According to the results in [32, Corollary 2.1], if $W_i(b_i^{n})$ 
is increasing with $b_i^{n}$, then the Whittle’s
index for any state $b_i^{n}$ is nothing but $W_i(b_i^{n})$. To that extent, we prove that $W_i(b_i^{n})$ is increasing with $b_i^{n}$. However, since $b_i^{n}$ is increasing with $n$ when $m=b$ (Lemma \ref{lem:MAoII_increasing}), it is sufficient to show that $W_i(b_i^{n})$ is increasing with $n$ to establish the desired result.  

Therefore, we first seek a closed-form expression of the intersection point $W_i(b_i^{n})$, we obtain:
\begin{align}
W_i(b_i^{n})=&\frac{(1-r_i)^2 \rho_i}{r_i^2}-\frac{(1-N_ir_i)^2 \rho_i}{(N_ir_i)^2} \nonumber \\ 
&+(1-N_ir_i)^{n+2}(n \rho_i+1+\frac{\rho_i(1-N_i r_i)}{N_i r_i}) \nonumber \\
& \ \ \ \times [\frac{1-(1-\rho_i)(1+(N_i-1)r_i)}{N_i r_i (1-(1-\rho_i)(1-r_i))} ]\nonumber \\
&-(1-r_i)^{n+2}(n \rho_i+1+\frac{\rho_i(1-r_i)}{r_i}) \nonumber \\
& \ \  \ \times [\frac{\rho_i}{ r_i (1-(1-\rho_i)(1-r_i))}]
\end{align}
Now, we provide the main result that allows us to affirm that $W_i(b_i^{n})$ is effectively the Whittle index of state $b_i^{n}$.
\begin{lemma}
The sequence $W_i(b_i^{n})$ is increasing with $n$.
\end{lemma}
\begin{IEEEproof}
After some mathematical analysis and algebraic manipulations, we get:
\begin{align}
W_i(b_i^{n+1})&-W_i(b_i^{n})=(n\rho_i+1)[(1-r_i)^{n+2}-(1-N_ir_i)^{n+2}] \nonumber \\
+&\frac{(n\rho_i+1)(1-\rho_i)}{1-(1-\rho_i)(1-r_i)}r_i[N_i(1-N_ir_i)^{n+2}-(1-r_i)^{n+2}]
\end{align}
We have that:
\begin{equation}
N_i(1-N_ir_i)^{n+2}-(1-r_i)^{n+2} \geq (1-N_ir_i)^{n+2}-(1-r_i)^{n+2}
\end{equation}
Thus:
\begin{align}
W_i(b_i^{n+1})-W_i(b_i^{n}) \geq &(n\rho_i+1) \times [1-\frac{(1-\rho_i)}{1-(1-\rho_i)(1-r_i)}r_i] \nonumber \\ &\times [(1-r_i)^{n+2}-(1-N_ir_i)^{n+2}]
\end{align}
Given that $(1-r_i)^{n+2}-(1-N_ir_i)^{n+2} \geq 0$ and $1-\frac{(1-\rho_i)}{1-(1-\rho_i)(1-r_i)}r_i=\frac{\rho_i}{1-(1-\rho_i)(1-r_i)} \geq 0$, therefore:
\begin{equation}
W_i(b_i^{n+1})-W_i(b_i^{n}) \geq 0
\end{equation}
That concludes the proof. 
\end{IEEEproof}
\end{appendices}

\end{document}